# Interplay between Temperature Oscillations and Melt Pool Dynamics in 3D Manufacturing Techniques

Stepan L. Lomaev[1,*], Georgii A. Gordeev[2], Marat A. Timirgazin[1], Dinara R. Fattalova[1] and Mikhail D. Krivilyov[1,2]

[1]Udmurt Federal Research Center of the Ural Branch of RAS, Baramzina str. 34, Izhevsk 426067, Russia
[2]Udmurt State University, Universitetskaya str. 1, Izhevsk 426034, Russia

**ABSTRACT**. The aim of this paper is coupling of temperature oscillations and melt pool dynamics experimentally observed in laser melting. The literature survey has shown that the developed explanations are mainly focused on the capillary and hydrodynamic aspects of the problem. As shown, complete analysis is only possible if the temperature oscillations are properly accounted for. Specifically, this effect is a governing mechanism that controls the melt pool dynamics via feedback coupling of heating, evaporation, capillarity and free-surface contraction. The present study suggests a physically consistent model validated on the detailed data of the time-resolved absorptance measured in [Phys. Rev. Appl. 10, 044061 (2018)]. An analytical equation is derived for oscillation spectra within this approach. It has been proved that the surface oscillations can be initiated without the keyhole effect. However, keyhole formation can cause additional modes to appear in the oscillation spectrum. The practical results include the formulas that are convenient for real-time monitoring of the surface temperature and characteristic flow velocity in the molten pool using the measured absorptance spectra. The impact of the temperature coefficient of surface tension on the maximal temperature, pool length, natural frequency of the free surface oscillations, and attenuation coefficient is thoroughly studied. All results are presented as closed-form formulas suitable for design of industrial laser systems.

## I. INTRODUCTION.

The laser processing of metals, including welding and additive 3D-printing, is becoming increasingly important in manufacturing parts and assemblies in various production areas, such as the automotive and aerospace industries [1]. In practical terms, the development, configuration and implementation of such methods require substantial time and production resources [2]. It occurs because the quality of parts fabricated with laser melting technologies deeply depends [3] on the dynamical stability of the melt pool (MP). High temperature gradients, multiple undergoing phenomena like the Marangoni effect, ablation vapor pressure, potential keyhole formation, and a phase transition between solids and liquids directly influence the complex hydrodynamic pattern formed in MP. An interaction between different driving forces and dissipation effects combined with the nonlinearity of thermophysical material properties leads to unsteady flow, deformation of the free surface and different oscillation processes [4,5]. The unstable behavior of MP in its turn affects the quality of a seam in laser welding or of a metal layer deposited in additive manufacturing. Finally, it may cause a production failure with the generation of geometrical and metallurgical defects [6].

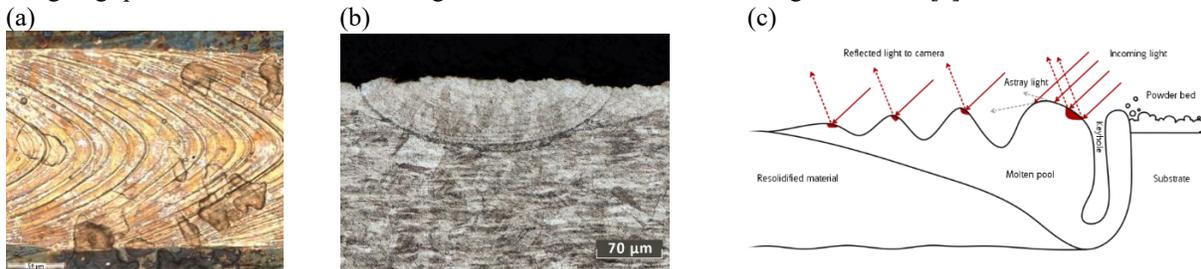

FIG 1. Experimental evidence of the effect of thermal oscillations in the melt pool (MP). (a) A chevron-type pattern observed at the solidified track after linear laser processing with remelting. (b) Periodic embowed strips registered at the metallographic cross section after chemical etching. (c) Experimental scheme for direct registration of capillary waves by the measurement of light reflections from the MP surface [7].

Oscillation of MP can be clearly detected experimentally based either on the as-processed samples and the results of in situ measurements. In Fig. 1, a typical representation of such oscillations is depicted. First, a characteristic chevron-type pattern is observed at the track's surface after laser skin remelting of the metal sample (Fig. 1(a)). The observed picture can be explained only by the oscillations of the melt surface during melting and subsequent solidification. Periodic embowed strips are often registered at the metallographic cross section (Fig. 1(b)). In some experiments [7], the capillary waves are photographed directly by measurement of light reflections over the MP surface (Fig. 1(c)).

*Contact author: lomaevst@udman.ru

Therefore, the discussed oscillations of MP may defer the solidification front and hence influence the final microstructure.

The oscillation effects in MP during laser processing have been studied for laser welding in the literature [8] where this phenomenon is explicitly linked to the seam quality. A complex recirculation pattern in the melt is established that changes the energy transport not only inside MP but also towards the solid metal matrix. As a result, such coupling impacts the MP shape [9,10,11]. The theoretical description of MP oscillations in welding is based on a dominated effect of keyhole formation, where the oscillation frequency depends on the geometrical parameters of the fluctuating keyhole cavity [12,13] and capillary force effects. However, the proposed model cannot explain the observed MP oscillations if keyholes are absent. In the review [14], different processing regimes are analyzed to reveal how a transition from deep-keyhole to keyhole-free occurs if the scanning speed increases and the MP oscillations are weakened. A qualitative criterion for the formation of highly oscillating stream patterns is proposed, which accounts for the inclination angle of the keyhole back wall.

In some studies, a possibility for quality optimization in laser processing is proposed using curvilinear trajectories in laser scanning [15]. As shown, if the laser beam moves along a circular or more complex trajectory like Lissajous figures, then the temperature profile is substantially modified and the MP oscillations are suppressed. It develops a tool for modulation of the MP shape, elimination of keyholes and microstructure control during solidification.

Recently, some attempts were made to use MP oscillations for real-time monitoring and diagnosis of ongoing laser processing. For example, in [7] the measured frequency of MP oscillations was used for detection and prediction of deep keyholes in selective laser melting. The authors [16] studied the correlation between the oscillations of the absorption coefficient at early stages of keyhole formation. The oscillations of thermal and concentration fields in the tail section of MP may affect the solidification process, leading to the formation of segregation strips also known as molten metal trails reported in [17]. To summarize, there are some perspectives on the practical application of the discovered effect in solidification control.

More recently, a new technique for microstructure modification has been developed based on in situ ultrasonic treatment of MP in laser processing methods [18,19,20]. In [20], the ultrasound was generated by optoacoustic effects via intense material evaporation from the melt surface. The effect of temperature oscillations analyzed in the present paper also leads to ultrasound generation through intensive evaporation of metal. Thus, it confirms the substantial application-oriented importance of research in this direction.

To summarize, we would like to emphasize that understanding MP oscillations holds promise both for the development of the techniques of monitoring the MP conditions and the strategies of controlling the crystallization process. There is a significant number of works devoted to modeling physical phenomena in MP [21,22,23,24]. However, studying the MP oscillations numerically is a difficult task, because this phenomenon is linked to complex interactions between different physical processes within a medium at different scales. Here, the problem of computational instability of the solutions becomes significant. For instance, in numerical models with free surface evaluation [22,23], it was found that surface oscillations can change the flow patterns and the MP shape. The response of the oscillation spectra of the free surface to different modulations of laser irradiation and temperature-dependent thermophysical properties was analyzed for keyhole-free processing in [24]. This numerical study showed that even small changes in free surface oscillations yield dramatic changes in temperature and velocity distributions, with further alteration of the MP shape.

Our interest lies in MP oscillation modes that do not significantly alter the MP shape or the pattern of hydrodynamic flows. Fig. 1(c) provides a schematic representation of this mode, and its effect on the melt zone can be seen in Fig. 1(a) and Fig. 1(b). Such moderate oscillations are of particular interest for this research, as they are of the only type that can be used for monitoring and control of the crystallization process. This article provides an in-depth analysis of the physical processes that lead to the oscillation effect in MP. It proposes a physically consistent model based on the perturbation theory, which has been cross-verified against experimental data. Within the framework of the model, it will be shown that the MP oscillations are not caused by the effect of keyhole formation as was previously reported [12,13]. The keyhole may be a contributing factor that influences the final spectrum of the MP oscillations, but is not a necessary one. As revealed below, the MP oscillations are caused by a combination of thermodynamic and hydrodynamic factors.

## II. FEEDBACK MODEL OF TEMPERATURE OSCILLATIONS AND MELT POOL DYNAMICS

Numerical simulation of the temperature oscillations is a complex computational task because it requires solution of coupled unsteady problems of laser energy absorption, heat transfer, multiphase hydrodynamics and phase transformations with latent heat at the metal surface. Owing to this reason, computer modeling does not produce accurate solutions of the mentioned coupled problems if the amplitude of the temperature oscillations is significant. In the approach suggested here, the search for solutions is split into two steps.

(1) All coupled problems are solved in the quasi-stationary approximation, assuming that the MP geometry, temperature, and velocity fields are stationary in a moving coordinate system associated with a moving laser beam.

(2) The MP oscillations are described using perturbation theory, where small deviations in temperature and related quantities reflect the unsteady behavior of the analyzed process. The feedback model is based on application of the small-perturbation theory to the heat transfer equation with convection

$$\dot{T} - \alpha \Delta T + \mathbf{v} \cdot \nabla T = \frac{q}{c_p \rho}, \qquad (1)$$

where temperature $T = T(x,y,z,t)$, flow velocity $\mathbf{v} = \mathbf{v}(x,y,z,t)$ and the heat source intensity $q = q(x,y,z,t)$ are the functions of time $t$ and coordinates $(x,y,z)$, $\alpha$ is the thermal diffusivity of material in MP, $c_p$ is the isobaric specific heat, $\nabla$ is the differential operator, $\dot{T} = \frac{\partial T(x,y,z,t)}{\partial t}$ is the time derivative of temperature. Further analysis of convection in MP is based on the Navier-Stokes equations for incompressible liquids:

$$\frac{\partial \mathbf{v}}{\partial t} = -(\mathbf{v} \cdot \nabla)\mathbf{v} + \nu \Delta \mathbf{v} - \frac{1}{\rho} \nabla P + \mathbf{g}, \qquad (2)$$

$$div\, \mathbf{v} = 0, \qquad (3)$$

where $P = P(x,y,z,t)$ is the pressure distribution formed either by capillary forces and reaction pressure owing to material evaporation in MP, $\mathbf{g}$ is the gravitational acceleration, $\nu = \frac{\eta}{\rho}$ is the kinematic viscosity, and $\Delta = \nabla^2$ is the Laplace operator.

In the general case, the heat transfer problem Eq. (1) must be solved in the 3D formulation. However, with small temperature perturbations $\delta T$ along the MP surface, it can be reduced to a one-dimensional approximation if $\delta T$ do not lead to a change in the MP shape and internal flow patterns. Here, Eqs. (1)–(3) after a series of transformations are turned into a single equation

$$\ddot{\delta T} + 2\zeta \omega_0 \dot{\delta T} + \omega_0^2 \delta T = f_{noise}(t), \qquad (4)$$

with the added noise function $f_{noise}(t)$ that includes random influences of various kinds. Expression (4) is the oscillator equation with attenuation, where the primary oscillating variable is the temperature deviation $\delta T$ from the mean value at the MP surface in the vicinity of the laser beam focus. Then, the natural frequency of the analyzed system is defined by

$$\nu_0 = \frac{\omega_0}{2\pi} = \frac{1}{2\pi l}\sqrt{\frac{\theta B P_{vap}}{\rho\, T_{max}}}, \qquad (5)$$

with the dimensionless attenuation coefficient

$$\zeta = \frac{2+\theta}{2\sqrt{\theta}} \frac{v_{ch}}{l\, \omega_0}, \qquad (6)$$

where $l$ is a characteristic distance between the maximum of MP surface temperature and MP boundary, $v_{ch}$ is the characteristic convective velocity in MP, $T_{max}$ is the maximum temperature at the MP surface defined at the laser beam center in the quasi-stationary approximation (i.e. without temperature oscillations), $P_{vap} = P_{vap}(T_{max})$ is the reactive pressure produced by evaporating substance at the MP surface, $B$ is the saturated vapor constant (see Eq. (7)), and $\theta$ is the dimensionless coefficient for correction of the temperature distribution in MP.

In the model Eqs. (4)–(6), MP is considered as a thermodynamic system with a following feedback scheme:

(1) A small increase in the maximum surface temperature $T_{max}$ leads to an increase in the reactive vapor pressure $P_{vap}$. Growth of $P_{vap}$ results in the enhancement of the velocity magnitude $|\mathbf{v}|$ in hydrodynamic flow patterns inside MP;

(2) Increase in $|\mathbf{v}|$ facilitates heat rejection at the MP surface according to Eq. (2). Better cooling yields decay of $T_{max}$;

(3) Then the feedback cycle is repeated for a small decrease of $T_{max}$. If the attenuation coefficient $\zeta$ in Eq. (6) is sufficiently small, for instance $\zeta \leq 0.25$, then the temperature deviation $\delta T$ starts oscillating with a frequency close to the natural frequency $\nu_0$.

From Eqs. (5) and (6), the attenuation coefficient $\zeta$ is defined by the temperature $T_{max}$, thermophysical characteristics of the melt, and is linearly proportional to the characteristic flow velocity $v_{ch}$. Thus, the parameter $v_{ch}$ acts as an amplifier of the dissipation factor for the oscillating system. Therefore, temperature oscillations are not generated in MP with intensive hydrodynamic mixing. Eq. (6) can now be used as a criterion of amplification or decay of the temperature oscillations.

Notably, the natural frequency $v_0$ does not depend on $v_{ch}$ directly. Hence, robust simulation of hydrodynamic flows is not required in the present study of temperature oscillations. This is a significant advantage of our approach. For proper analysis, the evaluation of characteristic values of $v_{ch}$ under the quasi-stationary approximation suffices to determine how fast the MP oscillations decay.

From Eq. (5), the frequency $v_0$ is determined by the distance $l$, maximum $T_{max}$ of the MP surface temperature and by material properties. In the present model, the vapor pressure $P_{vap}$ is calculated using the exponential function:

$$P_{vap} = \left(1 - \frac{\phi}{2}\right) P_0 e^{A - \frac{B}{T_{max}}}, \qquad (7)$$

where $P_0 = 1$ Pa is a normalization factor, $A$ and $B$ are the coefficients derived from material properties and obtained experimentally. The coefficient $\phi$ defines a fraction of particles (molecules) that does not return into MP after evaporation from its surface. In laser processing of metallic materials, the coefficient $\phi$ may vary [25] in the range $\phi \in [0; 0.82]$.

Eqs. (4)–(6) are derived independently for a specific temperature distribution inside MP. The equations remain correct for any continuous decreasing function of temperature $T(x)$, where the spatial coordinate $x$ takes the values from the MP center to the solidification front. In the limiting case, $T(x)$ takes the form of the Rosenthal solution and has the hyperbolic type if $v_{ch} \ll \sqrt{\frac{\alpha}{l}}$ [26]. Otherwise, if $v_{ch} \gg \sqrt{\frac{\alpha}{l}}$ the solution for $T(x)$ tends to be linear. Under processing conditions typical in selective laser melting (SLM) and directed energy deposition (DED), an intermediate thermal mode usually realizes. In [27], real-time measurements of temperature at the MP surface during DED are performed with the experimental data provided in Figure 17 of [27]. The surface temperature at the tail of MP is assumed linear. However, this approximation has a sufficient accuracy only in a localized MP domain. Over the whole MP surface, the temperature is accurately described by an exponential function as

$$T_s(x) = T_{max} \left(\frac{T_{liq}}{T_{max}}\right)^{\frac{x}{l}}, \qquad (8)$$

where $T_{liq}$ is the liquidus temperature of the material in MP. Based on the fitting given by Eq. (8), after a series of mathematical transformations of Eq. (1) towards Eq. (4), the following expression for the correction coefficient $\theta$ of temperature distribution from Eq. (6) has been received as

$$\theta = ln\left(\frac{T_{max}}{T_{liq}}\right). \qquad (9)$$

Next, laser treatment of metals is accompanied by multiple effects that reveal themselves as sources of thermal perturbations. These phenomena include light reflection on surface irregularities and powder particles, fluctuations of the irradiated power density, focus adjustment of the laser beam and others. The perturbations have a stochastic behavior and are encountered during the whole period of laser processing. In the present paper, the family of all mentioned effects is formalized by a noise function $f_{noise}(t)$ on the right-hand side of Eq. (4). Hence, Eq. (4) is mathematically classified as a differential functional, which converts the noise function into a frequency spectrum.

Eqs. (4)–(9) define the oscillations of the MP surface temperature $T_s$ unambiguously. The spectrum itself does not depend on the amplitude of oscillations and is determined by the maximum surface temperature $T_{max}$ averaged over time, MP width, material properties and noise function $f_{noise}$. Further verification of the equations is provided below.

Finally, we present another important theoretical implication of the formulated analytical model in this section. Eq. (5) defines the function $v_0(T_{max}, l)$. At the same time, the model allows to solve the inverse problem for determination of the maximal melt pool temperature as

$$T_{max}(v_0, l) \approx (1 + \lambda)\left(-\frac{1}{2}T_{boil}\left(2\ln(\gamma l v_0) - 1 + \sqrt{(2\ln(\gamma l v_0) - 1)^2 - \frac{4B}{T_{boil}}}\right)\right) - \lambda T_{boil}, \qquad (10)$$

where the coefficient $\gamma$ depends on thermophysical properties

$$\gamma = 2\pi \sqrt{\frac{\rho \, T_{boil}}{\left(1-\frac{\phi}{2}\right) P_0 e^{A_B} \ln\left(\frac{T_{boil}}{T_{liq}}\right)}}. \quad (11)$$

Eqs. (10) and (11) are derived in the approximation form and they precisely correspond to Eq. (5) only at $T_{max} = T_{boil}$. The correction factor $\lambda \approx -0.1$ is required for error compensation. It is independent on $l$, $\nu_0$ and is dependent on thermophysical material properties. For the parameters of the stainless 316L steel defined in Table I, the best correspondence between Eqs. (10), (11) and (5) is obtained at $\lambda = -0.11$. In this case, the calculation error does not exceed 5 K in the temperature range $T_{max} = T_{boil} \pm 300$ K and it diminishes as $T_{max}$ approaches $T_{boil}$. The obtained Eqs. (12) and (13) can be used to analyze the maximum temperature values on the melt surface $T_{max}$ in real time during or after laser t processing via analysis of the chevron-type patterns (see Fig. 1).

### III. ASSESSMENT OF FREE SURFACE OSCILLATIONS

Real-time in situ measurement of temperature oscillations at the MP surface is technically difficult because laser processing is complicated by intensive material evaporation and high instability of the MP surface. Temperature registration should be performed at a sampling frequency greater than the natural frequency. Therefore, verification of the formulated model is not performed by comparison with direct temperature measurements but using a set of experimental data on the time-dependent absorptance. The effect of temperature oscillations is directly linked to fluctuations of the reactive vapor pressure at the MP surface and hydrodynamic flow oscillations inside MP. Above a certain amplitude, these oscillations may cause a substantial deformation of the MP surface beneath the laser beam. Consequently, the laser absorptance shows conjugated fluctuations thanks to multiple reflections of laser irradiation. In [16], a comprehensive experimental study of the absorptance oscillations is completed for laser processing of a flat stainless 316L plate. The authors [16] registered the light reflected by MP using an integrating calorimetric sphere. All tests included the 10 ms accumulation period with the laser pulse energy in the interval $E_{in} \in [1.2; 6.3]$ J. The fulfilled analysis delivered the absorptance oscillations in real time with the time resolution of less than 1 μs. The obtained in [16] spectra are used for verification of the model presented by Eqs. (4)–(9).

The most distinct oscillations of absorptance are registered at the pulse energy of $E_{imp} = 3.1$ J and $E_{imp} = 3.27$ J (with the original notation $E_{in}$ in [16]). For these regimes, the authors of [16] deliver both the absorptance dynamics and frequency spectra after the fast Fourier transform. The highest peaks in the spectra correspond to the frequencies $\nu_{ex} = 8.6$ kHz and $\nu_{ex} = 6.4$ kHz correspondingly. In [16], the relevant data on the MP width are provided with $W_{weld} = 418$ μm and $W_{weld} = 468$ μm correspondingly. In Eq. (1), the parameter $l$ is defined as a distance from the MP center to its boundary Thus, the experimental half-width of MP can be used in calculations as

$$l_{exp} = \frac{1}{2} W_{weld}. \quad (12)$$

Table I shows the saturated vapor constants for 316L taken from [28] for calculation of the function $\nu_0(T_{max})$ defined by Eq. (5) which expresses the natural frequency $\nu_0$ via the maximal temperature $T_{max}$.

Fig. 2 depicts the calculated plots $\nu_0(T_{max})$. In all plots, the blue and orange lines are related to the processing modes No. 1 ($E_{imp} = 3.1$ J) and No. 2 ($E_{imp} = 3.27$ J) correspondingly. The experimental data [16] are marked by the color lines emphasized by black dashes. In Fig. 2, the frequencies $\nu_{exp}$ of the maximal peaks registered in the experimental spectra are also drawn by two horizontal lines. The dashed vertical lines show the auxiliary values of temperature that correspond to $T_{max} = 3211$ K and $T_{max} = 3129$ K for the regimes 1 and 2 correspondingly, where the predicted function $\nu_0(T_{max})$ crosses the experimental values of $\nu_{exp}$.

Following the obtained results, the natural frequency $\nu_0$ increases if $T_{max}$ grows. The orange line is located below the blue one because the semi-width $l_{exp}$ of MP in the regime No. 2 is greater than in the regime 1. The calculated $\nu_0$ should match the experimental frequencies $\nu_{exp}$, thus the maximum surface temperatures are defined from this coupling as $T_{max} = 3211$ K and $T_{max} = 3129$ K for the regimes No. 1 and 2 correspondingly.

TABLE I. Thermophysical properties of the 316L melts used in the calculations.

| Property | Notation | Value | Units |
|---|---|---|---|
| Saturated vapor constants | $A$ | 25.601 [28] | – |
|  | $B$ | $43.445 \times 10^3$ [28] | K |
| Density | $\rho$ | $6.2 \times 10^3$ [29] | kg/m³ |
| Viscosity | $\eta$ | $8.0 \times 10^{-3}$ [30] | Pa s |
| Boiling temperature at standard pressure | $T_{boil}$ | 3090 [28] | K |
| Liquidus temperature | $T_{liq}$ | 1660 [31] | K |
| Thermal conductivity | $k$ | 35 [31] | W/(m K) |
| Specific heat capacity | $c_p$ | 950 [29] | J/(kg K) |
| Thermal diffusivity | $\alpha$ | $0.59 \times 10^{-5}$ | m²/s |
| Specific heat of evaporation | $L_{vap}$ | $7.45 \times 10^6$ [28] | J/kg |
| Surface tension | $\sigma(T_{liq})$ | 1.5 [32] | N/m |
| Temperature coefficient of the surface tension | $\beta = d\sigma/dT$ | $\{-4; -2; -1; 0; 1; 1; 2; 4\} \times 10^{-4}$ | N/(m K) |

The authors [16] do not provide accurate numbers on the surface temperature $T_{max}$. Only the estimate $T_{max} \geq 3300$ K is given that is depicted in Fig. 2 by a green vertical line. At such temperatures, the calculated natural frequencies take the values $\nu_0 \geq 10.4$ kHz and $\nu_0 \geq 9.3$ kHz according to our model. Thus, one may conclude that the fair agreement with [16] in the frequency $\nu_0$ of thermal oscillations is achieved. After further analysis, this acceptable discrepancy of about 20% with [16] may come from an overestimated value $T_{max} \geq 3300$ K. In [16], the temperature evaluation is only based on the simulation results of spot welding [33] owing to technical difficulties in direct measurement. Since the simulated problem [33] does not fit the conditions of experiments [16], extended computer modeling of the studied problem has been performed in the present paper.

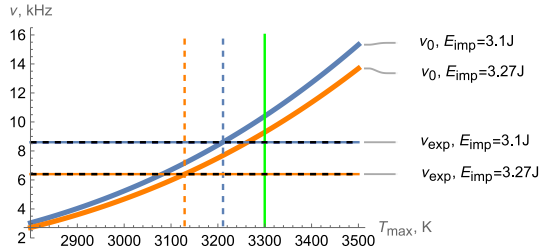

FIG 2. Cross-comparison of the analytical model $\nu_0(T_{max})$ (color solid curves) and experimentally detected frequencies $\nu_{exp}$ [16] (horizontal dashed lines at the frequencies of $\nu_{ex} = 8.6$ kHz and $\nu_{ex} = 6.4$ kHz) provided for the processing modes No 1. (pulse energy $E_{imp} = 3.1$ J, blue color) and No. 2 ($E_{imp} = 3.27$ J, orange color). The points of intersection between the model and experimental data provide the maximum temperatures at the melt pool surface predicted by Eqs. (5)–(9) at the values $T_{max} = 3211$ K and $T_{max} = 3129$ K. The vertical green line fits to the value $T_{max} = 3300$ K used in [16].

## IV. COUPLING FREE SURFACE OSCILLATIONS WITH CAPILLARY AND THERMOCAPILLARY FORCES

For numerical simulation of laser melting a plate surface by a stationary laser, a computer model [34] is implemented for description of conjugated heat transfer, laser surface annealing, fluid flow and free-surface dynamics with a moving boundary. The laser source is continuous and has top-hat spot at the focal plane that is 303 μm in diameter [16]. Secondary reflections of laser irradiation are also considered at the keyhole-type MP interface. Owing to the axial symmetry, the problem is solved within the 2D formulation using the cylindrical coordinate system. The mathematical model is resolved in Comsol MultiPhysics software with the finite element discretization. The free surface of MP is tracked using the arbitrary Euler-Lagrange method with a non-linearly deformed mesh.

For better computational stability, most thermophysical parameters of the melt in MP are selected constant at the temperature close to the boiling temperature $T_{boil}$, as has been done in [20]. A list of these thermophysical properties and references to original papers describing thermophysical measurements is presented in Table I. Two parameters including the surface tension $\sigma$ and its temperature coefficient $\beta = \frac{d\sigma}{dT}$ are selected as variables because of the following reasons. The hydrodynamic conditions inside MP influence its geometry and the maximal temperature $T_{max}$ at the MP surface. The latter value defines both the natural frequency $\nu_0$ and its attenuate coefficient $\zeta$ following Eqs. (5) and (6).

A quick assessment of the thermocapillary effect on fluid flow is conducted via the Marangoni number defined as

$$\text{Ma} = \left|\frac{d\sigma}{dT}\right| \frac{1}{\eta \alpha} l \, \Delta T \approx 6 \times 10^6 \left|\frac{d\sigma}{dT}\right|, \quad (13)$$

where the characteristic temperature difference $\Delta T = 1430$ K is equal to the difference between the boiling and melting temperatures at normal pressure. Other used parameters are taken from Table I. At $\text{Ma} > 10^3$, active thermocapillary convection is assumed, hence the hydrodynamic system is sensitive to the temperature coefficient $\beta$ if its deviations are about $\Delta \beta \approx 2 \times 10^{-4}$ N/(m K).

Analysis of the literature on experimental measurements of the surface tension [32,35,36,37,38] reveals that different research groups have obtained different values in the range $\sigma \in [1.3\,;1.9]$ N/m and for its temperature coefficient in the range $\beta \in [-4 \times 10^{-4}; 4 \times 10^{-4}]$ N/(m K). This difference in experimental data can be explained by several factors. First, measuring the surface tension and temperature coefficient is a challenging task, especially at high temperatures. Second, the temperature dependence of the temperature coefficient can vary significantly depending on the oxygen and sulfur content of the melt.

The provided description of the experimental method in [16], which was used as a basis for model verification, does not allow us to determine the values of $\sigma(T_{liq})$ and $\beta$ that best correspond with the experiment data.

Thus, we have performed a series of calculations based on a linear approximation of the temperature dependence $\sigma(T)$ of the surface tension. Each calculation corresponds to a specific value of the temperature coefficient of the surface tension $\beta = \{-4; -2; -1; 0; +1; +2; +4\} \times 10^{-4}$ N/(m K). The surface tension at the liquidus temperature have been chosen equal $\sigma(T_{liq}) = 1.5$ N/m as reported in [32].

The experimental data in Ref. [16] (illustrated by Figure 7 of [16]) evidence that the transition from a transient mode towards an oscillatory mode in MP occurs at $t \approx 2$ ms. At $t = 3$ ms, the absorptance oscillations achieve a steady behavior for the processing regimes with the pulse energies $E_{imp} = 3.1$ J и $E_{imp} = 3.27$ J. Our numerical simulations show the similar transition time $t = 3$ ms before the governing parameters $T_{max}$, $l$ and $v_{ch}$ achieve a quasi-static level with imposed oscillations. Therefore, in the post-simulation analysis all time-dependent functions of $T_{max}$, $l$ and $v_{ch}$ are averaged in time over the time interval $t \in [3; 4]$ ms. Before time averaging, the flow velocity $v_{ch}$ is first averaged spatially over the MP surface.

The results of computer simulations are summarized in Table II that also contains the natural frequency $v_0$ and attenuation coefficient $\zeta$ calculated with the analytical model presented by Eqs. (5) – (9). Fig. 3 visualizes the data from Table II as functions of the temperature coefficient $\beta$ of surface tension. The results calculated at $\beta = -4 \times 10^{-4}$ N/(m K) are not enough accurate because time averaging is conducted over the limited time intervals $t \in [3; 3.6]$ ms and $t \in [3; 3.3]$ ms for the processing regimes No. 1 and 2 correspondingly because of weak computational stability

Analysis of the maximal temperature $T_{max}$ (Fig. 3(a)) yields that its values are smaller than the estimate $T_{max} \geq 3300$ K given in [16]. We hypothesize it happens owing to two factors: (i) intensive convective heat transfer in MP, (ii) deep keyhole-like deformation of the MP surface. For further evaluation, the typical profiles, flow velocity distribution and streamlines in MP are illustrated in Fig. 4 for the processing regime No. 2 at different values of $\beta = \{-10^{-4}; 0; 10^{-4}\}$ N/(m K). The value and sign of $\beta$ have a great impact on hydrodynamics in MP (Fig. 4) and then on the semi-width $l$ (Fig. 3(b)) and maximal temperature $T_{max}$ (Fig. 3(a)). At negative $\beta$, the mass flux is directed from a hot MP center to its border, hence the semi-width $l$ increases rapidly (Fig. 4(a)) and MP deepens with the proceeded keyhole formation. Contrary, at positive $\beta$ the melt stream heads towards the hot MP central point resulting in a completely different flow pattern (Fig. 4(c)) with a single toroidal vortex. Consequently, MP has the smaller semi-width and volume. The liquid-gas interface is less bended and the liquid-solid interface is shrunk comparing to the case of $\beta < 0$. This feature explains why at positive $\beta$ the characteristic flow velocity $v_{ch}$ is almost 3 times larger at the MP surface. Additionally, at positive $\beta$ the heat input inside MP is reduced. It happens owing to the convexed MP surface that adsorbs less laser irradiation than the deeply concaved shape at $\beta < 0$ where the keyhole effect yields the secondary reflection and increases the heat input.

.

TABLE II. Summary of the melt pool characteristics obtained using direct computer simulation of underlying thermal and capillary phenomena (described by the parameters $T_{max}$, $l$ and $v_{ch}$) and analytical calculus of the temperature oscillations by Eqs. (5)–(9) (defined through $v_0$ and $\zeta$). In the simulated series, the variable parameter $\beta$ is the temperature coefficient of surface tension, $T_{max}$ is the maximal temperature at the melt pool surface, $l$ is the MP semi-width, $v_{ch}$ is the characteristic flow velocity in MP, $v_0$ is the natural frequency of the MP free surface, $\zeta$ is the attenuation coefficient. The data on $T_{max}$, $l$ and $v_{ch}$ are obtained by averaging over the time interval $t \in [3; 4]$ ms after a pulse start when the oscillation mode is already established. The data marked by * are averaged over a limited time interval

| | Processing mode No. 1 with the energy pulse $E_{imp} = 3.1$ J | | | | | |
|---|---|---|---|---|---|---|
| | $\beta$, N/(m K) | $T_{max}$, K | $l$, μm | $v_{ch}$, m/s | $v_0$, Hz | $\zeta$ |
| * | $-4 \times 10^{-4}$ | 3035 | 252 | 0.934 | 4744 | 0.208 |
| | $-2 \times 10^{-4}$ | 3043 | 249 | 0.810 | 4896 | 0.177 |
| | $-1 \times 10^{-4}$ | 3079 | 239 | 0.735 | 5566 | 0.146 |
| | $0 \times 10^{-4}$ | 3178 | 204 | 0.478 | 8198 | 0.075 |
| | $1 \times 10^{-4}$ | 3083 | 178 | 0.941 | 7545 | 0.186 |
| | $2 \times 10^{-4}$ | 3067 | 175 | 1.571 | 7385 | 0.327 |
| | $4 \times 10^{-4}$ | 3077 | 171 | 2.572 | 7742 | 0.515 |
| | experimental data from [16], $\beta$ is unspecified | $\geq 3300$ | 209 | unspecified | 8600 | unspecified |
| | Processing mode No. 2 with the energy pulse $E_{imp} = 3.27$ J | | | | | |
| | $\beta$, N/(m K) | $T_{max}$, K | $l$, μm | $v_{ch}$, m/s | $v_0$, Hz | $\zeta$ |
| * | $-4 \times 10^{-4}$ | 3056 | 266 | 1.189 | 4731 | 0.251 |
| | $-2 \times 10^{-4}$ | 3060 | 264 | 0.878 | 4814 | 0.183 |
| | $-1 \times 10^{-4}$ | 3070 | 257 | 0.733 | 5066 | 0.149 |
| | $0 \times 10^{-4}$ | 3128 | 231 | 0.650 | 6462 | 0.115 |
| | $1 \times 10^{-4}$ | 3097 | 179 | 0.934 | 7756 | 0.178 |
| | $2 \times 10^{-4}$ | 3083 | 176 | 1.627 | 7631 | 0.351 |
| | $4 \times 10^{-4}$ | 3092 | 172 | 2.628 | 7977 | 0.507 |
| | experimental data from [16], $\beta$ is unspecified | $\geq 3300$ | 234 | unspecified | 6400 | unspecified |

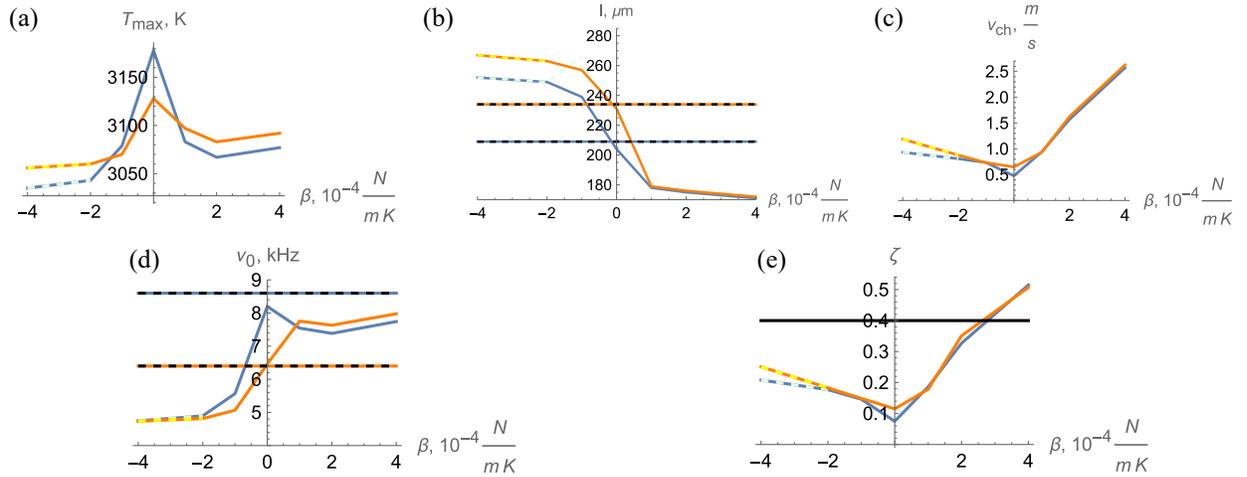

FIG 3. Physical characteristics of the melt pool (MP) and the oscillation dynamics of its surface as functions of the temperature coefficient $\beta$ of surface tension. (a) $T_{max}$, K; (b) $l$, μm; (c) $v_{ch}$, m/s; (d) $v_0$, Hz; (e) $\zeta$. The plotted characteristics have the same meaning and values as described in Table II. The blue and orange polylines correspond to the processing modes No. 1 ($E_{imp} = 3.1$ J) and No. 2 ($E_{imp} = 3.27$ J). The solid segments are obtained for time averaging in the interval $t \in [3; 4]$ ms while the dashed segments are received over the reduced intervals. The experimental data [16] are depicted by the color-and-dash horizontal lines.

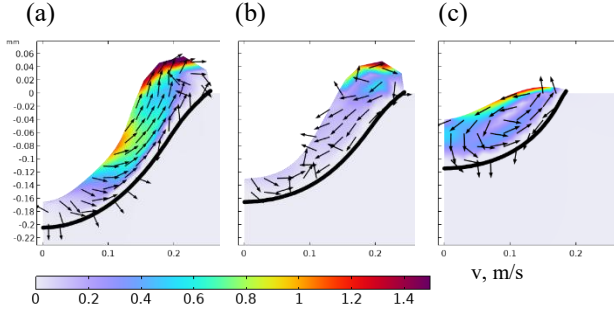

FIG 4. Shape of the melt pool, stream arrows and flow velocity magnitude in the azimuthal projection section calculated as a function of the temperature coefficient $\beta$ of surface tension in the quasi-stationary state. (a) $\beta = -10^{-4}$ N/(m K); (b) $\beta = 0$ N/(m K); (c) $\beta = 10^{-4}$ N/(m K). The parameters of the processing regime No. 2 with the pulse energy $E_{imp} = 3.27$ J are used. The solid black line corresponds to the liquidus isotherm and depicts the pool boundary.

The maximal values of $T_{max}$ (Fig. 3(a)) at both processing regimes are observed at $\beta = 0$ N/(m K) that is directly linked with the minimal value of the velocity $v_{ch}$ (Fig. 3(c)). The value $\beta = 0$ N/(m K) yields the best agreement (see Table II) to the experimental data for: (i) semi-width $l_{sim} = 204$ μm versus $l_{exp} = 209$ μm (regime No. 1); $l_{sim} = 231$ μm versus $l_{exp} = 234$ μm (regime No. 2); (ii) natural frequency $\nu_{0,sim} = 8198$ Hz versus $\nu_{0,exp} = 8600$ Hz (regime No. 1); $\nu_{0,sim} = 6462$ Hz versus $\nu_{0,exp} = 6400$ Hz (regime No. 2).

As shown, Eq. (5) either provides fair quantitative agreement to experimental data and predicts correctly a tendency of the natural frequency $\nu_0$ decay with the increase of the pulse energy $E_{imp}$. This tendency is logical because a higher pulse energy results in an increase of the semi-width of MP, hence a larger semi-length $l$ leads to reduction of $\nu_0$ according to Eq. (5). At the same time, the behavior of $T_{max}$ is also important. The maximal temperature decreases as $E_{imp}$ increases at $\beta = 0$ N/(m K)[1] (see Table II and Fig. 3(a)). The noticed coupling for $T_{max}$ with the processing regimes is unusual at the first glance and it requires explanations. Fig. 5 represents the temperature and flow velocity distributions in MP for two processing regimes at $t = 3.67$ ms.

The decay of $T_{max}$ in case of the $E_{imp}$ increase occurs owing to the combination of few interrelated factors:

(1) Increase of the flow velocity magnitude. Fig. 3(c) and Fig. 5 demonstrate the largest difference in the characteristic velocity magnitudes $v_{ch}$ for the different processing regimes is predicted at $\beta = 0$ N/(m K).

(2) Substantial increase of the MP bend and subsequently decrease of the distance $h$ between the maximal temperature point and the melting front ($h_1 = 663$ μm at regime No. 1 and $h_2 = 164$ μm at regime No. 2). A draft estimation $\frac{T_{max}-T_{liq}}{h}$ of the temperature gradient in the MP center gives $\frac{dT}{dz} \approx 2.3$ K/μm at the regime No. 1 contrary to $\frac{dT}{dz} \approx 8.8$ K/μm at the regime No. 2. Therefore, in spite of $T_{max}$ decay, the temperature gradient in MP grows by few times if the pulse energy variates.

(3) Increase of the MP volume and redistribution of temperature inside MP. A bigger semi-width leads to a larger surface area of MP. Using temperature distribution defined by Eq. (6) and the data from Table II, one may easily calculate the internal energy inside the MP surface layer within the cylindrical approximation. The internal energies divided by the layer thickness are $\approx 1606$ J/m and $\approx 1943$ J/m at the processing modes No. 1 and No. 2 correspondingly. Thus, the higher pulse energy, the larger energy is accumulated in the surface layer of MP even though the maximal energy is smaller.

In summary of the section, the verification procedure of Eq. (5) coupled to direct finite-element simulations has shown fair agreement of the calculated values with the experimental data [16] at two different processing regimes. As shown, the sign and magnitude of the temperature coefficient of surface tension $\beta$ have a great effect on the MP shape, convection intensity, maximal temperature, keyhole formation and even on the MP dynamics. The best agreement with experiments in the natural frequency and MP semi-length is got at $\beta = 0$ N/(m K).

It is important to note that, based on the results obtained, we cannot conclude that there is no thermocapillary effect in the experiment presented in [16]. The problem of precise determination of the temperature coefficient of surface tension will be highlighted further in Section VI.

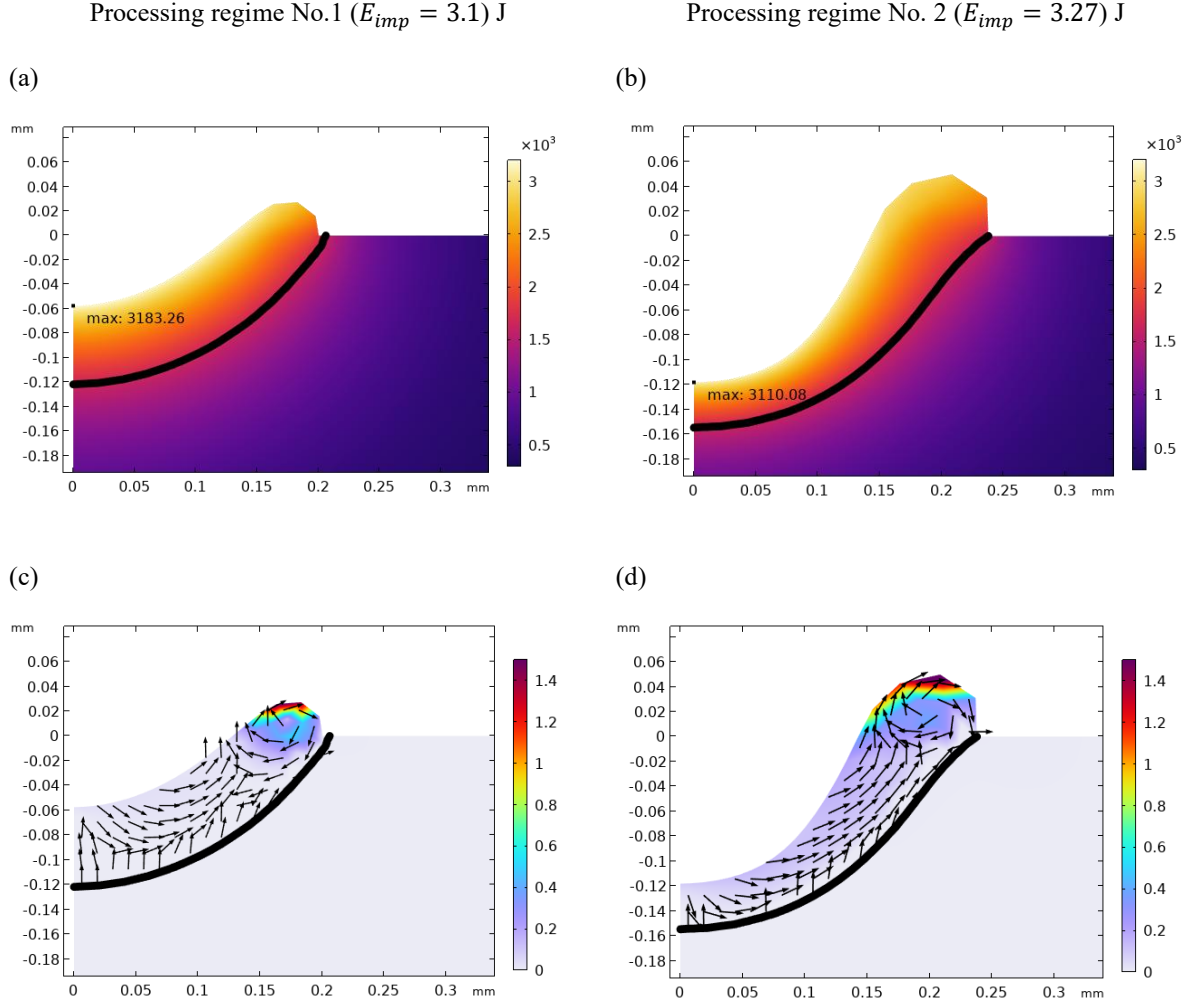

FIG 5. Temperature (a),(b) and flow velocity (c),(d) distributions in the melt pool at the processing regime No. 1 (subfigures (a),(c)) and regime No. 2 (subfigures (b),(d)) at the time $t = 3.67$ ms. The black line is the liquidus isotherm and the arrows show the flow pattern.

## V. FREQUENCY ANALYSIS

Verification of Eq. (5) in the previous section is built on comparison of the calculated natural frequency $\nu_{0,sim}$ and the maximal peak frequency $\nu_{0,exp}$ registered experimentally [16]. However, the value of $\nu_{0,exp}$ is not always equal to the natural frequency of oscillations and it depends on either the system configuration and external actuation that leads to temperature perturbations $\delta T(t)$ at the MP surface. In other words, determination of $\nu_{exp}$ is possible only via solution of the differential Eq. (4) which includes the external noise actuation $f_{noise}(t)$. Such effects are stochastic in their nature thus the function $f_{noise}(t)$ in the formulated model we consider as white noise. Its discrete analogue has been modeled with Wolfram Mathematica and then used for numerical solution of Eq. (4) as follows:

(1) An array $\{x_i\}$ of $10^4$ random numbers in the range of $[-1; 1]$ is first generated. Each number is associated to the time array $\{t_i\}$ with the period $\tau = 1$ µs as $t_i = t_0 + \tau \times i$. The value of $\tau$ is selected based on the sampling period of the experimental setup [16]. The total time interval is 10 ms and it is equal to the laser pulse duration in [16];

(2) Then the continuous function $f_{noise}(t)$ is constructed applying cubic interpolation splines. The derived function $f_{noise}(t)$ is used for numerical solution of Eq. (4) for determination of the temperature oscillation function $\delta T(t)$;

(3) After Fourier transformation, the frequency spectrum of $\delta T(t)$ is obtained.

The simulated results [39] and their match to experimental data [16] are summarized in Fig.6. The spectra are received at different processing regimes and temperature coefficients $\beta$ taken from Table II.

The differential Eq.(4) is linear. Hence, the calculated spectra and their maxima in Fig. 6(c-h) do not depend on the noise amplitude of $f_{noise}(t)$. The results are only sensitive to the probability distribution of random numbers used for construction of $f_{noise}(t)$. For each individual run, $f_{noise}(t)$ is freshly generated for better statistical confidence of $\nu_0$. In Fig. 6(a) and Fig. 6(b), the difference between $\nu_0$ and $\nu_{exp}$ is equal or even smaller than the one between $\nu_0$ and the frequency maximum in the calculated spectra.

The next step of the present analysis is devoted to a shape of the obtained spectra. In the suggested model, the temperature fluctuations $\delta T$ are considered as small. The dependent fluctuations of flow velocity $\delta v$ and reactive vapor pressure $\delta P$ are assumed as linearly dependent on $\delta T$ with sufficient accuracy. We imply that the oscillations of the absorptance coefficient can be attributed to $\delta T$ linearly as well. Therefore, a comparison of both the maximal frequencies and the spectrum profiles is logical, including the width of the resonance peak and level of background fluctuations. Both characteristics depend on the attenuation coefficient $\zeta$ which in its turn is linearly proportional to $v_{ch}$. Hence, a correspondence between spectrum profiles in [16] and calculated by Eq. (4) becomes an additional method of verification of temperature oscillations in MP.

Following Figs. 6 (c-h), the resonance peak goes wider and fuzzier if $\zeta$ increases. The background oscillations in the experimental spectrum at the processing mode No. 1 do not exceed the amplitude of 0.2 in Fig. 6(a). Although the spectrum for the processing regime 2 in Fig. 6(b) reveals few additional peaks at the frequencies of 9.6, 11.9 and 13.6 kHz whose origin is discussed in what follows. In Fig. 6(b), the background oscillations in the experimental spectrum for the regime No. 2 are below the level 0.3. The correlation analysis of the background signals and resonance peak widths between Figs. 6(a,b) and Figs. 6(c-h) demonstrates the best agreement with the experimental data in the case $\beta = 0$ N/(m K). Thus, it yields an extra confirmation of quantitative accuracy of the developed analytical model represented by Eqs. (4)–(9).

Concerning the additional minor peaks in Fig. 6(b), they are initiated by secondary modes in the frequency spectrum related to elasticity of the melt surface. Usually, this effect is prominent with well-developed keyhole formation. In the present model, it is not accounted for in the model, thus the calculated spectra do not contain the secondary peaks. Nevertheless, the model accurately predicts the principal mode of oscillations that is defined by the coupling between thermophysical and hydrodynamic phenomena in the melt pool.

The application of the external source term $f_{noise}(t)$ with white noise in Eq. (4), the resonant peak becomes imperceptible if the attenuation coefficient $\zeta \geq 0.4$ (see Fig. 3(e)). This conclusion is acceptable for any welding and 3D-printing processing regimes. Thus, Eq. (6) with Eq. (9) propose a new approach for prediction and control of the temperature oscillations in the melt pool.

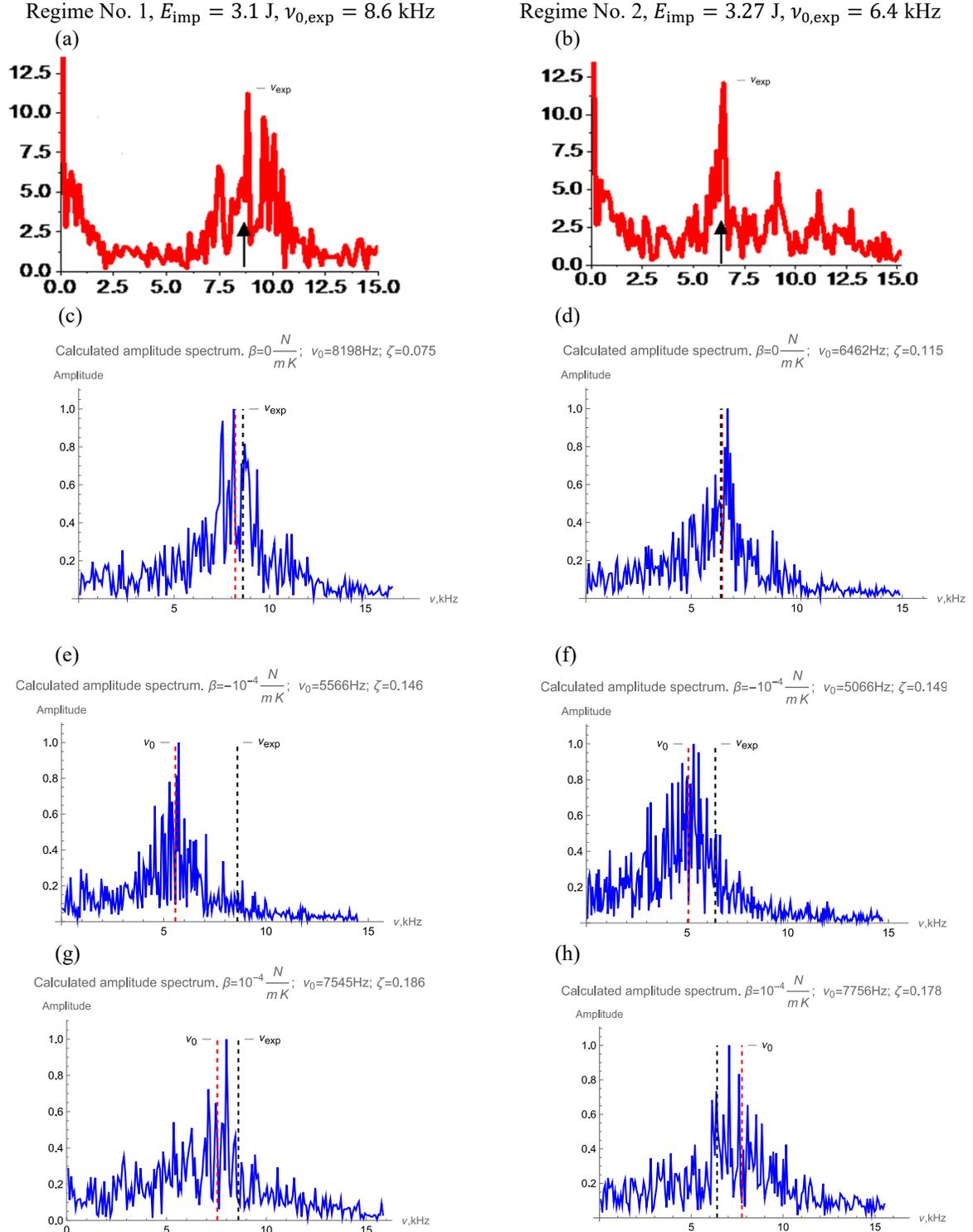

FIG 6. (a,b) Experimental frequency spectra of oscillations of the melt pool absorptance. Reprinted from Ref. [16], copyright 2018, with permission from APS. The vertical arrows mark the frequencies $v_{exp} = 8.6$ kHz and $v_{exp} = 6.4$ kHz with the maximal peak intensities (i.e. local maxima). The same values are depicted by dashed black lines in subfigures (c-f) for comparison with the calculated spectra [39]. (c,d,e,f) Simulated frequency spectra obtained after numerical solution of Eq. (4) using the parameters in Table II and the variable coefficient (c,d) $\beta = 0$ N/(m K), (e,f) $\beta = -10^{-4}$ N/(m K), (g,h) $\beta = 10^{-4}$ N/(m K). The red dashed lines show the calculated natural frequencies $v_0$.

## VI. DISCUSSION
This section focuses on two significant methodological issues for this study's development. The first issue concerns the selection of the thermal coefficient $\beta = 0$ N/(m K) and the interpretation of the calculation results presented in Section IV. The second issue relates to prospective direct experimental verification of the model presented.

The selection of the temperature coefficient $\beta = 0$ N/(m K) requires further justification. In studies dedicated to the simulation of the melt pool hydrodynamics in layer-by-layer synthesis techniques, different values of $\beta$ are suggested. For instance, $\beta < 0$ N/(m K) is assumed in [4,40,41], while [42] used $\beta > 0$ N/(m K) in calculations. If the concentrations of oxygen and sulfur in the 316L SS melt are known, the reasonable approach [43] for replicating real experimental conditions is to take $\beta$ as a function of temperature like presented in Fig. 7.

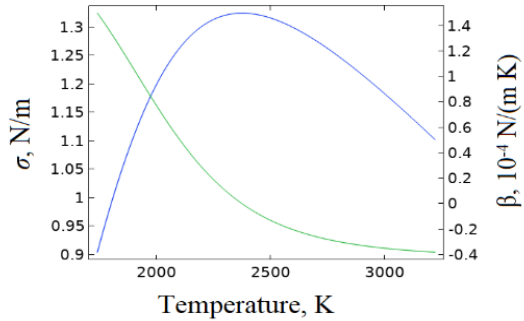

FIG 7. Surface tension $\sigma(T)$ and the temperature coefficient $\beta(T) = \frac{d\sigma(T)}{dT}$ of surface tension as functions of temperature depicted in blue and green correspondingly. The graphs illustrate the functional dependencies suggested in [43].

We cannot definitively state that the function shown in Figure 7 is the best fit for the experimental data presented in [16]. However, the type of dependence shown in Fig. 7 helps to interpret the selection of the temperature coefficient $\beta = 0$ N/(m K) when modeling proceeds within a linear approximation.

The function $\beta(T) = d\sigma/dT$ depicted in green in Fig. 7 is close to $\beta = 0$ N/(m K) in the temperature range $T \in [2100; 2400]$ K. These temperatures exceed the liquidus by between 600 and 900 K and are very typical for the melt pool surface. Therefore, $\beta = 0$ N/(m K) fits well with the thermophysical condition in the area of interest. Additionally, this value yields the best agreement between the calculated results and experimental data for three parameters simultaneously: (i) semi-width of the melt pool; (ii) natural frequency of surface oscillations, (iii) spectrum profiles which depend on the average convection level. Moreover, similar conclusions are drawn from the analysis of local hydrodynamic conditions in different zones of the melt pool. Its outermost and central zones have significant capillary forces and exhibits vortex flow structures. Both factors diminish the thermocapillary effect because of their opposite effect compared to Marangoni forces.

In conclusion, we consider $\beta = 0$ N/(m K) to be an effective value that provides the most accurate results within the linear approximation for the particular system and temperature range corresponding to the experiments presented in [16].

The model was verified using the data from measurements of laser radiation absorption in real-time [16]. One can believe that direct experimental verification of the model in which oscillations in temperature on the surface of a melt pool are registered will also be possible in the future. The direct experimental verification can be performed using in situ thermal monitoring of the melt pool that records the real-time temperature of the melt pool. Up to date, such monitoring systems employ the basic concept suggested by Concept Laser and the research group by J.P. Kruth in [44].

The control system combines two recording subsystems that complement each other. A single control device is incapable in this situation because of multiple reflection effects and instability of the measured signal. The CMOP camera is used for measuring the dimensions of the melt pool at frame rates between 5 and 10 kHz. Simultaneously, the high-temperature diode registers a radiation level emitted from the melt pool at even higher frequencies. The diode with a sampling rate above 20 kHz is sufficient for experimental verification of the effect of temperature oscillations suggested in our paper.

## VII. SUMMARY
This paper presents and validates a new analytical model of temperature oscillations applicable to laser welding and 3D-printing processes. The model employs perturbation theory and incorporates radiative and convective heat transfer, reactive vapor pressure and hydrodynamic effects. Based on the performed simulations and comprehensive comparison with experimental data, the following conclusions have been drawn.

(1) It has been proved that the origin of oscillations of the melt pool surface can be connected to temperature

fluctuations at the free surface caused by the feedback coupling between energy absorption, intensive metal evaporation and multiphase forced convection;

(2) The model developed in the present paper considers the effect of temperature oscillations as a primary (i.e. governing) mechanism that initiates periodic oscillations of the melt pool's surface. The keyhole effect is interpreted as a secondary (i.e. assistant) factor specifically in this matter. This assumption is made based on critical evaluation of available experimental data, including measured frequency spectra and the molten zone depth. Thus, the keyhole effect may give additional frequency peaks in the spectrum similar to the frequency modes depicted in Fig. 6b;

(3) The developed model introduces the analytic expression Eq. (5) for the natural (basic) frequency of $\nu_0$ that depends on the maximal temperature $T_{max}$ of the melt pool surface, its linear size $l$ and material thermophysical properties of the melt. The $\nu_0$ value increases as $T_{max}$ raises and it is inversely proportional to the $l$ value. The verification performed on reliable experimental data provided in literature [16] has shown the fair agreement between the calculated and measured $\nu_0$ in the absorptance spectra. The model also explains the registered tendency towards $\nu_0$ decrease if the pulse energy increases;

(4) Eq. (5) defines the function $\nu_0(T_{max}, l)$. The formulated analytical model of temperature oscillations allows to solve the inverse problem for determination of the maximal melt pool temperature through Eqs. (10) and (11). These equations are derived in the approximation form, and they correspond precisely to Eq. (5) only at $T_{max} = T_{boil}$. The correction factor $\lambda \approx -0.1$ is required for error compensation. It is independent on $l$, $\nu_0$ and depends on thermophysical material properties;

(5) For the parameters of the stainless 316L steel defined in Table I, the best correspondence between Eqs. (10), (11) and (5) is obtained at $\lambda = -0.11$. In this case, the calculation error does not exceed 5 K in the temperature range $T_{max} = T_{boil} \pm 300$ K and it diminishes as $T_{max}$ approaches $T_{boil}$. Therefore, Eqs. (10) and (11) are recommended for practical application for in-situ monitoring of the maximal melt pool temperature in real time. This method is more practical and accurate compared to absorptance measurements, where the melt pool temperature is substantially altered by intensive metal evaporation;

(6) An analytical expression Eq. (5) for the attenuation coefficient $\zeta$ is developed where $\zeta$ depends on $T_{max}$, characteristic flow velocity $v_{ch}$ and thermophysical properties of the melt. As follows from the performed analysis, $\zeta$ drops with $T_{max}$ increase and it is linearly proportional to $v_{ch}$. In case of $\zeta$ elevation, the resonance peak becomes wider and background fluctuations build up. Upon achieving the critical level $\zeta = 0.4$ the resonant peak disappears and the temperature oscillations decay. A semi-quantitative assessment has revealed a good match between the calculated and experimentally measured spectrum profiles;

(7) The thermocapillary (Marangoni) effect does not contribute any additional peaks into the oscillation spectrum. However, it may affect either $\nu_0$ and $\zeta$ indirectly via hydrodynamic coupling. In the present work, this effect is found significant. In the simulation series, the variation of the temperature coefficient $\beta$ of surface tension in the range $\beta \in [-4 \times 10^{-4}; 4 \times 10^{-4}]$ N/(m K) yields substantial deviations in the frequency response. The variations are in the ranges $\nu_0 \in [4744; 8198]$ Hz and $\nu_0 \in [4731; 7977]$ Hz at the processing regimes No. 1 (with the pulse energy $E_{imp} = 3.1$ J) and No. 2 ($E_{imp} = 3.27$ J). The attenuation coefficient fulfills the ranges $\zeta \in [0.075; 0.515]$ and $\zeta \in [0.115; 0.507]$ correspondingly, i.e. the dissipation rate changes drastically at different $\beta$ values;

(8) The best correspondence to the experimental data [16] is received for the semi-width ($l_{exp} = 209$ μm versus $l_{sim} = 204$ μm and $l_{exp} = 234$ μm versus $l_{sim} = 231$ μm at the processing regimes 1 and 2), natural frequency ($\nu_{exp} = 8600$ Hz versus $\nu_{0,sim} = 8198$ Hz, $\nu_{exp} = 6400$ Hz versus $\nu_{0,sim} = 6462$ Hz at the processing regimes 1 and 2) and spectrum profiles at $\beta = 0$ N/(m K). The value $\beta = 0$ N/(m K) is considered to be an effective value of the temperature coefficient of surface tension, which provides the most accurate fit within the linear approximation for the particular system and temperature range corresponding to the experimental data presented in [16];

(9) The case of $\beta = 0$ N/(m K) is particularly interesting because it shows an unusual decrease of the maximal temperature $T_{max}$ if the pulse energy increases from $E_{imp} = 3.1$ J to $E_{imp} = 3.27$ J. The physical reason for that behavior is a complex coupling between thermophysical and hydrodynamic phenomena under laser processing of a metal surface by a stationary pulse laser. Proper accounting of this effect may be important in modeling of 3D-printing of thin-wall parts where pulse lasers are usually utilized [45,46].

## ACKNOWLEDGMENTS
The work of S.L.L. and D.R.F. was carried out within the framework of the state tasks of the Ministry of Science and Higher Education of the Russian Federation, reg. No FUUE- 2024-0011.

The work of M.A.T. and M.D.K. was carried out within the framework of the state tasks of the Ministry of Science and Higher Education of the Russian Federation, reg. No FUUE-2024-0006.

G.A.G. acknowledges the support from the Russian Science Foundation grant (project No. 25-79-10169).

## AUTHOR CONTRIBUTIONS STATEMENT
The idea of the study was proposed by S.L.L. and G.A.G. S.L.L. conceptualized the study, performed the mathematical models, and wrote the manuscript. G.A.G. contributed to the design of the technique for simulations. M.A.T. performed a formal analysis of the mathematical models and simulated data. D.R.F. implemented the computer code and performed simulations. M.D.K. aided with conceptualization of the study and compilation of the manuscript. All authors have contributed to the discussion and approved the final version of the manuscript.

The authors declare no competing financial interest.

[1] R. Bayat, Z. Wang, T. Sparks, F. Liou and J. Newkirk, *Aerospace applications of laser additive manufacturing,* in *Laser Additive Manufacturing* edited by M. Brandt (Elsevier, 2017), pp. 351-371.

[2] D. D. Gu, W. Meiners, K. Wissenbach, and R. Poprawe, Laser additive manufacturing of metallic components: materials, processes and mechanisms, Int. Mater. Rev **57**, 133 (2012).

[3] G. Tapia and A. Elwany, A review on process monitoring and control in metal-based additive manufacturing, J. Manuf. Sci. Eng. **136**, 060801 (2014).

[4] S.A. Khairallah, A.T. Anderson, A. Rubenchik and W.E. King, Laser powder-bed fusion additive manufacturing: Physics of complex melt flow and formation mechanisms of pores, spatter, and denudation zones, Acta Mater. **108**, 36 (2016).

[5] C. Zhao, N.D. Parab, X. Li, K. Fezzaa, W. Tan, A.D. Rollett and T. Sun, Critical instability at moving keyhole tip generates porosity in laser melting, Science **370**, 1080 (2020).

[6] T. DebRoy and S. David, Physical processes in fusion welding, Rev. Mod. Phys. **67**, 85 (1995).

[7] L. Caprio, A. G. Demir and B. Previtali, Observing molten pool surface oscillations during keyhole processing in laser powder bed fusion as a novel method to estimate the penetration depth, Addit. Manuf. **36**, 101470 (2020)

[8] J. F. Lancaster, *The Physics of Welding*: International series on materials science of technology (Pergamon Press, New York, 1986).

[9] G. M. Oreper and J. Szekely, Heat- and fluid-flow phenomena in weld pools. J. Fluid Mech. **147**, 53 (1984).

[10] K. C. Mills and B. J. Keene, Factors affecting variable weld penetration, Int. Mater. Rev. **35**, 185 (1990).

[11] S. A. David, T. DebRoy and J. M. Vitek, Phenomenological modeling of fusion welding processes, MRS Bull. **19**, 29 (1994).

[12] H. Maruo and Y. Hirata, Natural frequency and oscillation modes of weld pools. 1st Report: Weld pool oscillation in full penetration welding of thin plate, Weld. Int. **7**, 614 (1993).

[13] C. D. Sorensen and T. W. Eagar, Modeling of oscillations in partially penetrated weld pools, J. Dyn. Syst. Meas. Contr. **112**, 469 (1990).

[14] R. Fabbro, Physical mechanisms controlling keyhole and melt pool dynamics during laser welding, Advances in Laser Materials Processing: Technology, Research and Application, Woodhead Publishing, 211 (2010).

[15] T. Girerd, A. Gameros, M. Simonelli, A. Norton and A.T. Clare, Modulation of melt pool behaviour using novel laser beam oscillation methods, J. Mater. Process. Technol. **325**, 118300 (2024).

[16] B. J. Simonds, J. Sowards, J. Hadler, E. Pfeif, B. Wilthan, J. Tanner, C. Harris, P. Williams, and J. Lehman, Time-resolved absorptance and melt pool dynamics during intense laser irradiation of a metal, Phys. Rev. Appl **10**, 044061 (2018).

[17] N. Yang, J. Yee, B. Zheng, K. Gaiser, T. Reynolds, L. Clemon, W. Y. Lu, J. M. Schoenung and E. J. Lavernia, Process-structure-property relationships for 316L stainless steel fabricated by additive manufacturing and its implication for component engineering, J. Therm. Spray Technol. **26**, 610 (2017).

[18] C. J. Todaro, M. A. Easton, D. Qiu, D. Zhang, M. J. Bermingham, E. W. Lui, M. Brandt, D. H. StJohn and M. Qian., Grain structure control

during metal 3D printing by high-intensity ultrasound, Nat. Commun. **11**, 142 (2020).

[19] Ivan A. Ivanov, Vladimir S. Dub, Alexander A. Karabutov, Elena B. Cherepetskaya, Anton S. Bychkov, Igor A. Kudinov, Artem A. Gapeev, Mikhail D. Krivilyov, Nikolay N. Simakov, S. A. Gruzd, S. L. Lomaev, V. V. Dremov, P. V. Chirkov, R. M. Kichigin, A. V. Karavaev, M. Yu. Anufriev and K. E. Kuper, Effect of laser-induced ultrasound treatment on material structure in laser surface treatment for selective laser melting applications, Sci. Rep. **11**, 23501 (2021).

[20] S.L. Lomaev, D.R. Fattalova, G.A. Gordeev, M.A. Timirgazin and M.D. Krivilyov, Quantitative efficiency of optoacoustic ultrasonic treatment in SLM, DED, and LBW applications, Sci. Rep. **15**, 1887 (2025).

[21] P. S. Cook and A. B. Murphy, Simulation of melt pool behaviour during additive manufacturing: Underlying physics and progress, Addit. Manuf. **31**, 100909 (2020).

[22] R. T. C. Choo, J. Szekely and R. C. Westhoff, Modeling of high-current arcs with emphasis on free surface phenomena in the weld pool, Welding J. **69**, 346 (1990).

[23] E. J. Ha and W. S. Kim, A study of low-power density laser welding process with evolution of free surface, Int. J. Heat Fluid Flow **26**, 613 (2005).

[24] A. Ebrahimi, C. R. Kleijn and I. M. Richardson, Numerical study of molten metal melt pool behaviour during conduction-mode laser spot melting, J. Phys. D: Appl. Phys. **54**, 105304 (2020).

[25] A. Klassen, T. Scharowsky and C. Körner, Evaporation model for beam based additive manufacturing using free surface lattice Boltzmann methods, J. Phys. D: Appl. Phys. **47**, 275303 (2014).

[26] D. Rosenthal, The Theory of Moving Sources of Heat and Its Application to Metal Treatments, Transactions of the American Society of Mechanical Engineers **68**, 849 (1946).

[27] C. Hao, Z. Liu, H. Xie, K. Zhao and S. Liu, Real-time measurement method of melt pool temperature in the directed energy deposition process, Appl. Therm. Eng. **177**, 115475 (2020).

[28] V. Bobkov, L. Fokin, E. Petrov, V. Popov, V. Rumiantsev and A. Savvatimsky, Thermophysical properties of materials for nuclear engineering: a tutorial and collection of data. IAEA, Vienna 18–21 (2008).

[29] P. Pichler, B. Simonds, J. Sowards and G. Pottlacher, Measurements of thermophysical properties of solid and liquid NIST SRM 316L stainless steel, J. Mater. Sci. **55**, 4081-4093 (2020).

[30] J. J. Valencia and P. N. Quested, Thermophysical properties. Metals process simulation, 18-32 (2010)

[31] K. C. Mills, *Recommended values of thermophysical properties for selected commercial alloys*. Woodhead publishing (2002).

[32] S. Ozawa, K. Morohoshi and T. Hibiya, Influence of oxygen partial pressure on surface tension of molten type 304 and 316 stainless steels measured by oscillating droplet method using electromagnetic levitation, ISIJ Int. **54**, 2097 (2014).

[33] W. Tan, S. B. Neil, and C. S. Yung, Investigation of key hole plume and molten pool based on a three-dimensional dynamic model with sharp interface formulation, J. Phys. D: Appl. Phys. **46**, 055501 (2013).

[34] D. R. Fattalova, G. A. Gordeev and S. L. Lomaev, Investigation of the pulsed laser parameters for optimizing SLM processes in the manufacture of metal products with improved characteristics, Bulletin of Perm University. Physics **2**, 47 (2025).

[35] H. Fukuyama, H. Higashi and H. Yamano, Thermophysical properties of molten stainless steel containing 5 mass% B4C, Nucl. Technol., 1 (2019).

[36] T. Dubberstein, H.P. Heller, J. Klostermann, R. Schwarze and J. Brillo, Surface tension and density data for Fe–Cr–Mo, Fe–Cr–Ni, and Fe–Cr–Mn–Ni steels, J. Mater. Sci. **50**, 7227 (2015).

[37] K. C. Mills, Y. Su, Z. Li and R. F. Brooks, Equations for the calculation of the thermophysical properties of stainless steel, ISIJ Int. **44**, 1661 (2004).

[38] J. Volpp, Surface Tension Estimation of Steel above Boiling Temperature, Applied Sciences **14**, 3778 (2024).

[39] Lomaev, S. (2025). The program code for calculating the data. https://doi.org/10.5281/zenodo.17466748

[40] L. Liu, M. Huang, Y. H. Ma, M. L. Qin and T. T. Liu, Simulation of powder packing and thermo-fluid dynamic of 316L stainless steel by selective laser melting, Int. J. Heat Mass Transfer **29**, 7369-7381 (2020).

[41] M. Bayat, V. K. Nadimpalli, D. B. Pedersen and J. H. Hattel, A fundamental investigation of thermo-capillarity in laser powder bed fusion of metals and alloys, Int. J. Heat Mass Transfer **166**, 120766 (2021).

[42] J. C. Haley, J. M. Schoenung and E. J. Lavernia, Modelling particle impact on the melt pool and wettability effects in laser directed energy deposition additive manufacturing, Mater. Sci. Eng., A **761**, 138052 (2019).

[43] C. Tang, K.Q. Le and C.H. Wong, Physics of humping formation in laser powder bed fusion, Int. J. Heat Mass Transfer **149**, 119172 (2020).

[44] S. Berumen, F. Bechmann, S. Lindner, J.-P. Kruth and T. Craeghs, Quality control of laser- and powder bed-based Additive Manufacturing (AM) technologies, Physics procedia, **5** 617-622 (2010)

[45] E. Vasileska, L. Caprio, A.G. Demir, B. Previtali and V. Gecevska, Defect classification and dimensional estimation in single-point exposure laser powder bed fusion of lattice struts using optical emission spectroscopy, Prog. Addit. Manuf., 1 (2025).

[46] A. D. Plessis, N. Razavi, M. Benedetti, S. Murchio, M. Leary, M. Watson, D. Bhate and F. Berto, Properties and applications of additively manufactured metallic cellular materials: a review, Prog. Mater Sci. **125**, 100918 (2022).